\magnification=\magstep1
\advance\vsize 0.1truein % so I'll fit on 4 pages.
\input vanilla.sty %%%%
\input pictex 
=%cmbx10 scaled\magstep1
    cmbx12                 
\font\small=cmr9
\font\smit=cmti9
\font\smsl=cmsl9    
\font\smmit=cmmi9
\font\smbf=cmbx9
\font\ninesy=cmsy9

\def\smalltype{\baselineskip=10pt \small
\textfont1=\smmit \textfont2=\ninesy 
\textfont\rmfam=\small
\textfont\bffam=\smbf \textfont\itfam=\smit \textfont\slfam=\smsl
\setbox\strutbox=\hbox{\vrule height7.5pt depth2.5pt width0pt}
\def\rm{\fam0\small}
\def\it{\fam\itfam\smit}
\def\sl{\fam\slfam\smsl}
\def\bf{\fam\bffam\smbf}
}

 \def\zT#1{{\text{#1}}}  
 \def\zZ#1{\ifcase#1 {}\or \displaystyle \or \textstyle
        \or \scriptstyle \or \scriptscriptstyle \fi}   
\def\ket#1{|#1\rangle}

{\advance\baselineskip1\jot         
\title \medbold A Possible New Quantum Algorithm: \\
\medbold
Arithmetic with Large Integers via the Chinese Remainder Theorem
\endheading}

\author S. A. Fulling \\
\rm Mathematics Department, Texas A\&M University,
College Station, TX 77843-3368
\endauthor

{\narrower\smalltype
Residue arithmetic is an elegant and convenient way of computing
with integers that exceed the natural word size of a computer.
The algorithms are highly parallel and hence naturally
adapted to quantum computation.
The process differs from most quantum algorithms currently 
under discussion in that the output would presumably be  
obtained by classical superposition of the output of many 
identical quantum systems, instead of by arranging for
constructive interference in the wave function of a single
quantum computer.
\par}\medskip 

Quantum computation is still a solution in search of problems.
To be profitably implemented on a (hypothetical) quantum computer, an
algorithm presumably must satisfy two requirements:
\smallskip
\item{1.} It must be parallelizable, so that identical operations are to
be performed simultaneously on different terms of a superposition state
vector.
\smallskip
\item{2.} There must be a way of reading out the answer by an observation
on the system --- i.e., on the final state vector as a whole.
Presumably this requires some kind of constructive interference of 
the terms in the superposition. 
 \smallskip 
 \noindent The second requirement 
is what distinguishes a quantum-computable problem from a generic 
parallelizable problem. 

 In the most intensively studied applications of quantum 
computing,
 factoring [1] and search [2], 
 most of the quantum operations are devoted to building up an
 eigenstate (of the computer's basic observables) that represents 
the result of the computation.
 The required Hilbert-space maneuvering gives each application an 
air of specialness.
 The present paper proposes a quantum algorithm of a different 
kind,
 in which the constructive interference that builds up the answer 
takes place at the level of classical waves or signals.
 Arguably, eventual general-purpose quantum computation is more 
likely to be of this type.

Arithmetic by the Chinese remainder theorem 
 is a highly parallel procedure [3--6].
The idea is that the fixed word size of a standard computer can be 
transcended by doing all addition, subtraction, and multiplication 
modulo each of a set of pairwise relatively prime integers 
(moduli), simultaneously. 
 The string of integers (each less than its 
corresponding modulus) uniquely represents some nonnegative integer 
less than the product of all the moduli.  
 Unlike standard digital 
arithmetic, the computations relative to each modulus are 
completely independent of each other (no carries or cross terms), 
so the process is immediately and completely  parallelizable. 

The algorithm for reconstructing the large integer in 
 human-readable (decimal) notation is relatively complicated and 
should be avoided or postponed whenever possible. 
 If the number is truly 
huge, even printing it out or transmitting it over a communications 
channel might be expensive, 
 and the usefulness of its exact 
expression open to question. 
 Instead, the advised strategy (in 
classical as well as quantum computation) is to do all computations 
with the large number internally, until a useful conclusion is 
reached. 
 This conclusion may be qualitative, such as that the 
number is exactly zero. 
 (For example, in quantum field theory the 
ratio of two such numbers may be the coefficient of some tensorial 
invariant in an effective Lagrangian. 
 The theory may be 
renormalizable only if that particular term is actually absent.) 
Or, it may suffice to know the number in floating-point form, to a 
precision much less than the total number of digits. 

We consider here some aspects of how  Chinese-remainder
arithmetic can be adapted to a quantum computer.
The obvious strategy is to divide the qubits of the computer into three
groups (data fields),
one to hold the modulus, one to hold the result of a computation modulo
that modulus, and one for work space (if needed).
The initial state should be a sum over moduli of states wherein the first
field holds the modulus and the other two are set to zero:
$$ \ket{\zT i} = \sum_{j=1}^R \ket{m_j}\ket{\bold0}\ket{\bold0}.$$
The computation should then read the modulus and perform the calculation
relative to that modulus, producing
 $$ \ket{\zT f\,} = \sum_{j=1}^R \ket{m_j}\ket{u_j}\ket{\bold0}.$$
(The notation is that of [4] and [6].)
We shall not discuss here how this initial state is to be prepared or how
the calculation is to be performed;
neither of these steps seems likely to be a major obstacle.
 (Quantum algorithms for the operations of modular arithmetic 
already appear in the literature --- e.g., [7].)

The interesting and novel aspect of this quantum algorithm is the output procedure.
We propose that the quantum apparatus can be made to emit a signal
(optical or electronic) that is (for the $j$th pure state) 
a sequence of
pulses with period $m_j\,$;
a pulse occurs when (and only when) the time is (a basic time unit times)
$u_j$ mod $m_j\,$.
 The examined output signal will be the
superposition of the signal from each term.
Instead of arranging for constructive interference in the state of the
qubits themselves and then measuring the qubits, therefore,
one will look for constructive interference in the output signal.
The strongest pulse will occur at time $u$, the number (less than
$\prod m_j$) that is congruent to $u_j$ mod $m_j$ --- 
 that is, the answer!

 The natural setting for this kind of procedure is a large ensemble 
of identical quantum computers.
 For example, in existing nuclear magnetic resonance experiments,
 each molecule  is a computer.
During ``readout'' each molecule independently collapses
into some eigenstate, with corresponding signal emission.
If more than one eigenstate is consistent with the computational
algorithm, therefore,
 the output from the whole system is
 an expectation value of the basic observable,
not an eigenvalue [8].
 Thus in such a computation
at least some of the superposition or interference takes 
place at the classical level, after the quantum observations have 
been performed.
 As explained in [8],
 although this feature is a nuisance in some  computations
 (such as quantum search when more than one object satisfies the 
search criterion),
 it is quite desirable in others, where the sought information is a 
statistical property of a quantum state rather than the random
 outcome of a single measurement on it.
 The present situation is akin to the latter.

Before discussing  practical difficulties of the proposed 
computation, 
  let us look at an example 
 (of course, with moduli too small and few to be of
ultimate interest).
Let $(m_1,m_2,m_3) = (3,4,7)$;
this is essentially the circle-of-fifths example in [6],
whose table the reader may find useful.
The numbers represented range from 0 to 83.
If $u=42$, then $(u_1,u_2,u_3) = (0,0,2)$.
The component of the signal from modulus 3 consists of pulses at
$t=0,3,6,\ldots$;
that from modulus 4 of pulses at $t=2,6,10,\ldots$;
that from modulus 7 of pulses at $t=0,7,14,\ldots$.
The total signal (shown in Figure~1)
has height 3 at $t=42$,
height 2 at 12 places, and height 1 at 36 places.
The pattern has a nontrivial structure, which is symmetric about 42.
(The pulses of height 2 fall into three classes, with respective
periodicities 12, 21, and 28 --- the pairwise products of the $m_i\,$.) 

\midinsert\centerline{\beginpicture
\setcoordinatesystem units <1cm,1.4mm>
\setplotarea x from -1 to 3, y from -84 to 0
\putrule from 0 0 to 2 0
\putrule from 0 -2 to 1 -2 
\putrule from 0 -3 to 1 -3 
\putrule from 0 -6 to 2 -6 
\putrule from 0 -7 to 1 -7
\putrule from 0 -9 to 1 -9 
\putrule from 0 -10 to 1 -10 
\putrule from 0 -12 to 1 -12 
\putrule from 0 -14 to 2 -14
\putrule from 0 -15 to 1 -15 
\putrule from 0 -18 to 2 -18 
\putrule from 0 -21 to 2 -21 
\putrule from 0 -22 to 1 -22 
\putrule from 0 -24 to 1 -24 
\putrule from 0 -26 to 1 -26 
\putrule from 0 -27 to 1 -27 
\putrule from 0 -28 to 1 -28 
\putrule from 0 -30 to 2 -30 
\putrule from 0 -33 to 1 -33 
\putrule from 0 -34 to 1 -34 
\putrule from 0 -35 to 1 -35 
\putrule from 0 -36 to 1 -36 
\putrule from 0 -38 to 1 -38 
\putrule from 0 -39 to 1 -39 
\putrule from 0 -42 to 3 -42 
\putrule from 0 -45 to 1 -45 
\putrule from 0 -46 to 1 -46 
\putrule from 0 -48 to 1 -48
\putrule from 0 -49 to 1 -49 
\putrule from 0 -50 to 1 -50 
\putrule from 0 -51 to 1 -51 
\putrule from 0 -54 to 2 -54 
\putrule from 0 -56 to 1 -56 
\putrule from 0 -57 to 1 -57 
\putrule from 0 -58 to 1 -58 
\putrule from 0 -60 to 1 -60 
\putrule from 0 -62 to 1 -62
\putrule from 0 -63 to 2 -63 
\putrule from 0 -66 to 2 -66
\putrule from 0 -69 to 1 -69 
\putrule from 0 -70 to 2 -70
\putrule from 0 -72 to 1 -72 
\putrule from 0 -74 to 1 -74 
\putrule from 0 -75 to 1 -75 
\putrule from 0 -77 to 1 -77 
\putrule from 0 -78 to 2 -78
\putrule from 0 -81 to 1 -81 
\putrule from 0 -82 to 1 -82
\putrule from 0 -84 to 2 -84
\put{$\zZ3 0$} [r] at -.2 0
\put{$\zZ3 4 $} [r] at -.2 -4
\put{$\zZ3 8 $} [r] at -.2 -8
\put{$\zZ3 12 $} [r] at -.2 -12
\put{$\zZ3 16 $} [r] at -.2 -16
\put{$\zZ3 20 $} [r] at -.2 -20
\put{$\zZ3 24 $} [r] at -.2 -24
\put{$\zZ3 28 $} [r] at -.2 -28
\put{$\zZ3 32 $} [r] at -.2 -32
\put{$\zZ3 36 $} [r] at -.2 -36
\put{$\zZ3 40 $} [r] at -.2 -40
\put{$\zZ3 44 $} [r] at -.2 -44
\put{$\zZ3 48 $} [r] at -.2 -48
\put{$\zZ3 52 $} [r] at -.2 -52
\put{$\zZ3 56 $} [r] at -.2 -56
\put{$\zZ3 60 $} [r] at -.2 -60
\put{$\zZ3 64 $} [r] at -.2 -64
\put{$\zZ3 68 $} [r] at -.2 -68
\put{$\zZ3 72 $} [r] at -.2 -72
\put{$\zZ3 76 $} [r] at -.2 -76
\put{$\zZ3 80 $} [r] at -.2 -80
\put{$\zZ3 84 $} [r] at -.2 -84  
\endpicture} \medskip
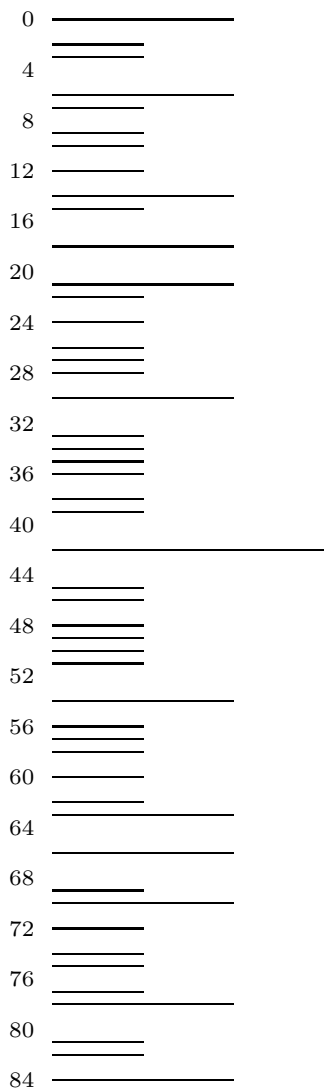
 \centerline{\small Fig.~1}\endinsert

There are two immediately obvious practical worries about this scheme.
First,  the bit pattern representing a number has length equal to the
largest integer represented ({\sl not\/} the number of digits in that
integer, which is logarithmically smaller).
By hypothesis, this number is very large.
Second, if $R$, the number of moduli, is not very small, it may be hard
to discriminate the pulse of height $R$ representing the answer from the
numerous pulses of height $R-1$.
Fortunately, the pulse pattern for a given sequence of moduli is unique,
and only its placement on the axis varies from one output number to
another;
furthermore, the pattern has such a structure that its center can still be
located even if there is a slight uncertainty in some of the pulse
heights.
Therefore,  to recognize the output number it is not necessary to
recognize, or even to wait for, the highest pulse;
one needs only to analyze enough of the output to recognize what part of
the pattern it is coming from.
(Implementation details obviously will require more input from number
theory, pattern recognition, and experimental expertise.)

This observation, however, raises another concern:  Since the 
pattern is always the same, the {\sl phase\/} of the output must be 
measured with absolute precision in order to determine the answer 
exactly.  
 But as previously remarked, quantum computation is no worse off 
than classical computation in this regard.  
 The indicated experimental procedure should be able to recognize 
the answer to a decent number of digits of precision, relative to 
the size of the product of moduli.  
   If the result of the computation is a qualitative 
conclusion that can be used internally by another kind of quantum 
computation, then the numerical readout procedure is not necessary 
at all. 

 It is difficult to be quantitative at this time about what 
capabilities for signal generation and detection would be necessary 
for a practical implementation of the algorithm:
 It is impossible to predict, for the time when serious quantum 
computers become reality,
 what will be the standard word size of a classical computer or 
the maximum size of an integer in a worthwhile computation.

 \smallskip\goodbreak
 {\bf Acknowledgments.}
 I thank Michael Kash, Andreas Klappenecker, and Davin Potts
 for helpful 
discussions, and Gordon Chen and Marlan Scully for stimulating my 
interest in quantum computation.

 \bigskip\goodbreak

 \centerline{\smc References}\medskip

  \frenchspacing\advance\parskip2\jot

 \item{1.} P. W. Shor, {\sl SIAM J. Comput. \bf26},
1484 (1997).

 \item{2.} L. K. Grover, {\sl Phys. Rev. Lett. \bf79},
 325 (1997).  

 \item{3.} I. Borosh and A. S. Fraenkel, {\sl Math. Comput. \bf 20},
 107 (1966).

\item{4.} D. E. Knuth, {\sl The Art of Computer Programming}, Vol. 2,
 {\sl Seminumerical Algorithms}, 2nd ed. (Addison--Wesley, Reading, 
1981), Sec. 4.3.2.

\item{5.} S. Szabo and R. Tanaka, 
 {\sl Residue Arithmetic and Its Applications to Computer 
Technology\/} (McGraw--Hill, New York, 1967).

  \item{6.} S. A. Fulling,  Large numbers, the 
Chinese remainder theorem, and the circle of fifths,
 submitted to {\sl Math. Mag.} 
 ({\tt http://www.math.tamu.edu/\~{}fulling/chinese.ps}).

  \item{7.} V. Vedral, A. Barenco, and A. Ekert,
{\sl Phys. Rev. A\/ \bf54}, 147 (1996).

 \item{8.} J. A. Jones and M. Mosca,
 {\sl Phys. Rev. Lett. \bf 83}, 1050 (1999).

 \bye